\documentclass[a4paper,10pt]{article}
\usepackage[onehalfspacing]{setspace}
\usepackage{amsmath}
\usepackage{amssymb}
\usepackage{hyperref}
\usepackage{fontenc}
\usepackage{inputenc}
\usepackage{authblk}
\usepackage{color}
\usepackage[top=1.5cm,bottom=2.5cm,right=2.54cm,left=2.54cm]{geometry}

\begin{document}

\title{\textbf{Weyl-invariant extension of the Metric-Affine Gravity}}
\author[1]{R. Vazirian\thanks{r.vazirian@srbiau.ac.ir}}
\author[2]{M. R. Tanhayi\thanks{m$_{-}$tanhayi@iauctb.ac.ir}}
\author[3]{Z. A. Motahar}
\affil[1]{\emph{\normalsize {Plasma Physics Research Center, Science
and Research Branch, Islamic Azad University, Tehran, Iran}}}
 \affil[2]{\emph{\normalsize{Department of Physics, Islamic Azad University, Central Tehran Branch, Tehran,
 Iran}}}
 \affil[3]{\emph{\normalsize{Department of Physics, University of Malaya, 50603 Kuala Lumpur, Malaysia }}}

\renewcommand\Authands{ and }

\maketitle

\begin{abstract}

Metric-affine geometry provides a non-trivial extension of the
general relativity where the metric and connection are treated as
the two independent fundamental quantities in constructing the
space-time (with non-vanishing torsion and non-metricity). In this
paper we study the generic form of action in this formalism, and
then construct the Weyl-invariant version of this theory. It is
shown that in Weitzenb\"{o}ck space, the obtained Weyl-invariant
action can cover the conformally invariant teleparallel action.
Finally the related field equations are obtained in the general
case.
\end{abstract}


\section{Introduction}

Extended theories of gravity  have become a field of interest in
recent years due to the lack of a theory which could fully describe
gravity as one of the fundamental interactions together with other
ones. Shortcoming of the current gravitation theory in cosmological
scales, despite all the successes in solar system tests, is another
reason in developing the idea of extending it and has confirmed the
need to surpass Einstein's general theory of relativity (GR)
\cite{1, 2}. The extension can be made in different ways such as
geometrical, dynamical and higher dimensional or even a combination
of them \cite{3,3.1}. In this paper, we will focus on geometric
extension to gravity under the notion of metric-affine formalism.

\smallskip
Our current description of gravity is based, via theory of GR, on
Riemannian geometry in which metric is the only geometrical object
needed to determine the space-time structure and connection is
considered to be metric-connection, the well known Christoffel
symbols. In such a context, as \emph{a priori}, the connection is
symmetric \textit{i.e.} torsion-free condition and is also assumed
to be compatible with the metric (${\nabla _\lambda }{g_{\mu \nu }} =
0$). To enlarge this scheme one can set the two presumptions aside
and think of metric-affine geometry in which metric and connection
are independent geometrical quantities. Therefore, connection is no
longer compatible with the metric and torsion-free condition is
relaxed, thus in addition to Christoffel
symbols, the affine connection would contain an anti-symmetric
part and non-metric terms as well. The transition from Riemann to metric-affine geometry has
led to Metric-Affine Gravity (MAG), an extended theory of gravity
which has been extensively studied during recent decades from
different viewpoints and a variety of non-Riemannian cosmology
models are proposed based on it (for example see \cite{hehl}).

\smallskip
The underlying idea of GR that explains gravity in terms of
geometric properties of space-time remains unaffected in MAG and
just new concepts are added to this picture by introducing the
torsion and non-metricity tensors in description
of gravity. Almost all of the space-time geometries such
as Riemann-Cartan, Weyl, Minkowski, etc. can be obtained by
constraining the three aforementioned tensors \cite{5, 11}; this is
one of the peculiarities of metric-affine formalism.

In this paper we study the conformally invariant metric-affine theory.
The conformal theories of gravity are in
great importance, for example, in the framework of quantum gravity
it is proved that such theories are better to re-normalize \cite{stelle}
and also from the phenomenological point of view it is shown that
the solutions of  conformal invariant theories (\textit{e.g.} Weyl
gravity) can explain the extraordinary speed of rotation curves of
galaxies and also they can address the cosmological constant problem
\cite{mann}. Thus considering the conformal invariance within
the context of MAG seems to be worthwhile which has been studied in
some papers \cite{luca, turk}. We first consider the action
in its general format in MAG and then fix the degrees of freedom by
imposing the conformal invariance.

The organization of the paper is as follows: the conventions and
general aspects of metric-affine geometry are briefly reviewed in
section 2. In section 3, the most general second order action in
metric-affine formalism is presented in terms of free parameters.
Weyl-invariant version of the action is investigated in section 4 by
determining the free parameters and the subject will be concluded in
the fifth section. The field equations of the general action are
presented in appendix.


\section{Metric-affine geometry}

As mentioned the main idea in MAG is the independence of metric and
connection, both of which being the fundamental quantities
indicating the space-time structure and carry their own dynamics in
contrast with GR where metric is the only independent dynamical
variable \cite{13, 14}. Note that in Palatini formalism of GR such an
assumption holds, but it brings nothing novel to the theory. In
presence of the gravitational field, space-time can be curved or
twisted and these features of space-time are characterized by
curvature and torsion tensors defined in terms of the
affine connection and its derivatives as
\begin{equation}
{\textbf{R}^\lambda }_{\beta \mu \nu } =  - {\partial _\nu }{\Gamma
^\lambda }_{\beta \mu } + {\partial _\mu }{\Gamma ^\lambda }_{\beta
\nu } + {\Gamma ^\lambda }_{\alpha \mu }{\Gamma ^\alpha }_{\beta \nu
} - {\Gamma ^\lambda }_{\alpha \nu }{\Gamma ^\alpha }_{\beta \mu }
\end{equation}
and\footnote{Square brackets and parenthesis in relations show
anti-symmetrization and symmetrization over indices.}
\begin{equation}
{S_{\mu \nu }}^\alpha  \equiv {\Gamma ^\alpha }_{[\mu \nu ]} =
\frac{{{\Gamma ^\alpha }_{\mu \nu } - {\Gamma ^\alpha }_{\nu \mu }}}
{2}{\text{ }}{\text{.}}
\end{equation}
These two definitions are obtained from the covariant derivatives
commutator acting on a vector field,
\begin{equation}
\left[ {{\nabla _\mu },{\nabla _\nu }} \right]{V^\lambda } =
{R^\lambda }_{\beta \mu \nu }{V^\beta } + 2{S_{\mu \nu }}^\alpha
{\nabla _\alpha }{V^\lambda }
\end{equation}
where $\nabla $ denotes covariant derivative associated with the
affine connection and for a tensor field is defined as
\begin{equation}
{\nabla _\mu }{A^\lambda }_\nu  = {\partial _\mu }{A^\lambda }_\nu +
{\Gamma ^\lambda }_{\beta \mu }{A^\beta }_\nu  - {\Gamma ^\beta
}_{\nu \mu }{A^\lambda }_\beta.
\end{equation}
It must be emphasized that the third index of the connection is
conventionally chosen to be in the direction along which
differentiation is done\footnote{The order of indices must be
carefully considered since the connection coefficients are supposed
not to be completely symmetric with non-vanishing torsion tensor.}
\cite{13, 15}.

\smallskip
Non-metricity tensor is the other geometrical object defined as
\begin{equation}
{\nabla _\lambda }{g_{\mu \nu }} \equiv {Q_{\lambda \mu \nu }}
\end{equation}
from which it is implied that inner product of vectors and so their
length and the angle between them are not conserved when parallel
transported along a curve in space-time \cite{3,16}.

\smallskip
Another difference arisen in this set-up is the increase of
dynamical degrees of freedom. Noting that the affine connection is
independent of the metric and is not constrained in general, thus it
contains 64 independent components in four-dimensional space-time.
As it is obvious by definitions given above, torsion and
non-metricity are rank-three tensors being anti-symmetric in first
and symmetric in last pair of indices respectively. Due to these
symmetry properties the two quantities will constitute the 64
independent components of the affine connection (24 components of
torsion tensor and 40 components of non-metriciy tensor). Thereby
the overall number of the independent components will become 74 by
taking into account the 10 independent components of the metric -- a
symmetric second rank tensor -- together with the connection.

\smallskip
It is straightforward to show that the affine connection can be
expressed in the following form
\begin{equation}
{\Gamma ^\lambda }_{\mu \nu } = \left\{ {{{^\lambda }_{\mu \nu }}}
\right\} + {K^\lambda }_{\mu \nu } + {L^\lambda }_{\mu \nu }
\label{affine connection}
\end{equation}
where $\left\{ {{{^\lambda }_{\mu \nu }}} \right\}$ is the usual
Christoffel symbol and the rest is a combination of torsion and
non-metricity tensors defined as follows
\begin{gather}
{K^\lambda }_{\mu \nu } = {g^{\sigma \lambda }}\left( {{S_{\mu
\sigma \nu }} + {S_{\nu \sigma \mu }} + {S_{\mu \nu \sigma }}}
\right), \notag\\
  {L^\lambda }_{\mu \nu } = \frac{1}
{2}{g^{\sigma \lambda }}\left( { - {Q_{\mu \nu \sigma }} - {Q_{\nu
\mu \sigma }} + {Q_{\sigma \mu \nu }}} \right). \label{connection
terms}
\end{gather}
Accordingly, the affine curvature tensor will take the form of
\begin{align}
  {\textbf{R}^\lambda }_{\beta \mu \nu } = {R^\lambda }_{\beta \mu \nu }(\{ \} )
  &- {\nabla _{\left\{ \nu  \right\}}}{L^\lambda }_{\beta \mu } + {\nabla _{\left\{ \mu  \right\}}}{L^\lambda }_{\beta \nu }
   + {L^\lambda }_{\alpha \mu }{L^\alpha }_{\beta \nu } - {L^\lambda }_{\alpha \nu }{L^\alpha }_{\beta \mu }
   \notag\\
   &- {\nabla _{\left\{ \nu  \right\}}}{K^\lambda }_{\beta \mu }
   + {\nabla _{\left\{ \mu  \right\}}}{K^\lambda }_{\beta \nu }
   + {K^\lambda }_{\alpha \mu }{K^\alpha }_{\beta \nu } - {K^\lambda }_{\alpha \nu }{K^\alpha }_{\beta \mu } \notag\\
   &+ {L^\lambda }_{\alpha \mu }{K^\alpha }_{\beta \nu } + {L^\alpha }_{\beta \nu }{K^\lambda }_{\alpha \mu }
    - {L^\alpha }_{\beta \mu }{K^\lambda }_{\alpha \nu } - {L^\lambda }_{\alpha \nu }{K^\alpha }_{\beta \mu }
\label{affine curvature}
\end{align}
where ${R^\lambda }_{\beta \mu \nu }(\{ \} )$ indicates Riemann
curvature tensor and braces are used to show covariant derivative in
Riemann geometry. Contraction of (\ref{affine curvature}) will
result in two different rank-two tensors \cite{3, 13}; one of
them is the affine Ricci tensor obtained by contracting the first
index with the last pair of indices (${\textbf{R}_{\beta \nu }}
\equiv {\textbf{R}^\lambda }_{\beta \lambda \nu }$) and the other
one is the so-called homothetic curvature tensor which is resulted
from contraction of the first two indices (${{\hat
{\textbf{R}}}_{\mu \nu }} \equiv {{\hat {\textbf{R}}}^\lambda
}_{\lambda \mu \nu }$). Generation of the two second rank curvature
tensors is due to the fact that the only symmetry of the affine
curvature tensor is the anti-symmetric property of the last pair of
indices. However, there is just one independent affine Ricci scalar
and the contraction of homothetic curvature tensor with metric gives
rise to vanishing scalar because of its anti-symmetric nature
\cite{3, 13}. Having more complete description of geometry, in
the following section we will focus on constructing the generic
gravitational action based on metric-affine geometry.


\section{Metric-affine second order action}

With independence of the metric and connection in MAG, the general
gravitational action is expressed in terms of metric, connection and
their derivatives. Also the coupling of matter action with
connection in addition to metric and matter fields is allowed in the
case of MAG which forms the dissimilarity between this approach and
the Palatini \cite{13, 17}. In this paper we limit our discussion to
the geometrical part of the action, but what the form of the action
is. Obviously, the first choice is to replace the Einstein-Hilbert
(EH) action with its counterpart in metric-affine formalism, but
this is not the only possibility and extension of EH action into a
more general one through the metric-affine formalism can be done by
considering torsion and non-metricity tensors as well as curvature
tensor in construction of the action \cite{note}. In order to
construct the gravitational Lagrangian, we follow the approach of
\cite{13} and \cite{19} which is an effective field theory approach
and appropriate scalar terms are constructed at each order by
applying power counting analysis. To start with and in natural units
$c=\hslash=1$, by choosing ${\text{[}}dx{\text{] = [}}dt{\text{] =
[}}l{\text{]}}$ all of the geometrical quantities which are needed
in construction of the action, are expressed in terms of length
dimension and, consequently, to make the action dimensionless, the
coupling constant is related to Planck length $l_p$. The highest
power of the length dimension in the scalars shows the order of
action so the first term of the generalized action which is the
Ricci scalar that is replaced by the affine one, is of second order.

\smallskip
Restricting our discussion to second order action and considering
symmetries of torsion and non-metricity tensors, there will be four
scalars written in terms of torsion tensor and its derivative\footnote{Terms of higher orders have been presented in \cite{10}
and for a less general case in \cite{13}.}, eight terms made up of
the non-metricity tensor and its derivative and three terms
constructed from the contraction of torsion and non-metricity
tensors. Accordingly, the generic second order
gravitational action in $n$ dimensions takes the following form\footnote{A similar action has been studied in
\cite{10} with different approach in the context of scalar, vector,
tensor theory. Note also that, it is proved that due to the symmetry properties, the other possible forms of non-vanishing scalars can be rewritten using the terms of this action.}
\begin{align}
 {\cal I}_{MAG} = \frac{1}
{\kappa}\int &{{{d^n}x} \sqrt { - g} \left( {{a_0}{\textbf{R}}}
\right.
 + {a_1}} {\nabla ^\mu}S_{\mu}
 + {a_2}S^{\mu}S_{\mu}
 + {a_3}S^{\mu}\,_ \lambda \,^\sigma {S_{\mu \sigma
 }}^\lambda + {a_4}S^{\mu \nu \lambda} S_{\mu \nu \lambda} \notag\\
 & + {a_5}{\nabla ^\mu }{Q^\sigma }_{\mu \sigma }
 + {a_6}{g^{\nu \sigma }}{\nabla _\mu }{Q^\mu }_{\nu \sigma }+ {a_7}{g^{\mu \nu }}{Q^\lambda }_{\mu \lambda }
 {Q^\sigma }_{\nu \sigma }
 + {a_8}{g^{\mu \lambda }}{Q^\nu }_{\mu \lambda }{Q^\sigma }_{\nu \sigma } \notag\\
 & + {a_9}{g^{\mu \nu }}{Q^\sigma }_{\mu \lambda }{Q^\lambda }_{\nu \sigma
 }+ {a_{10}}{g^{\mu \alpha }}{g^{\nu \beta }}{g_{\lambda \gamma }}{Q^\lambda }_{\mu \nu }{Q^\gamma }_{\alpha \beta }
 + {a_{11}}{g^{\mu \nu }}{g^{\alpha \beta }}{g_{\lambda \gamma }}{Q^\lambda }_{\mu \nu }{Q^\gamma }_{\alpha \beta }\notag\\
 &+ {a_{12}}{g^{\mu \nu }}{Q^\lambda }_{\mu \nu }{S_{\lambda}}
 + {a_{13}}{g^{\mu \nu }}{Q^\lambda }_{\mu \lambda }{S_{\nu}}
 \left. {+{a_{14}}{g^{\mu \alpha }}{Q^\lambda }_{\mu \nu }{S_{\lambda \alpha }}^\nu } \right) \label{Mag action}
\end{align}
in which ${S_\mu } \equiv {S_{\mu \beta }}^\beta$ and
${a_{i}}{\text{'s}}$ denote different coupling constants, $g$ stands
for the determinant of metric, and $\kappa$ contains the Planck
length, e.g. in four dimensions $\kappa=16\pi l_p^2$. The relevant
field equations can be obtained by varying the action with respect
to metric and connection independently, and then one obtains after a
trivial but rather lengthy calculation the field equations which are given in Appendix \href{App. A}A.\\
In the following section we are going to fix the free parameters
which appear in the action due to its general form of definition.


\section{Conformal invariance of the action}

The action in the form of (\ref{Mag action}) has free parameters
that must be fixed. This can be done by imposing some constraints or
initial conditions where, we use the conformal invariance condition. The first extension of Einstein's gravity was
done by Weyl in 1919 and then developed by Cartan and Dirac (for
review see \cite{ 2, 21}). In \cite{OBUKHOV}, conformal invariance
was considered in Riemann-Cartan geometry where the torsion plays
the role of an effective Weyl gauge field. Conformal torsion
gravity and conformal symmetry in teleparallelism were studied in \cite{Kazuharu}.  \\
Under the conformal transformation of metric of the form
\begin{equation}
{\overline g _{\mu \nu }}\left( x \right) = {\Omega ^2}{g_{\mu \nu
}}\left( x \right)\label{conformal metric},
\end{equation}
the non-metricity which is associated with $\nabla$, transforms as
\begin{equation}
{\bar Q^\lambda }_{\mu \nu } = 2{g_{\mu \nu }}{\nabla ^\lambda }\ln
\Omega  + {Q^\lambda }_{\mu \nu }\label{conformal non-metricity}
\end{equation}
where $\Omega $ is a scalar function of $x$. However, the conformal
transformation of the torsion tensor is somehow ambiguous and
different forms of transformation are considered; namely,
Weak conformal transformation under which the torsion tensor remains
unchanged and Strong conformal transformation in which the torsion
tensor transforms similar to (\ref{conformal metric}), namely one
has: ${\overline S _{\mu \nu }}^\lambda  = {\Omega ^2}{S_{\mu \nu
}}^\lambda $ \cite{luca}.  Being interested in conformal invariance
of the action, we choose the weak form which preserves the
anti-symmetric part of the affine connection. In accordance with the all above
mentioned, one can easily show that the affine connection is
not changed under the conformal transformation of the form
(\ref{conformal metric}).

\smallskip
It is straightforward to show that under conformal transformation,
the action (\ref{Mag action}) transforms in the following way:
\begin{align}
 {\overline {\cal I}_{MAG} } = {\Omega ^{n - 2}}{{\cal I}_{MAG}} + \frac{1} {\kappa}&\int
{{d^n}x}{\Omega ^{n - 2}}\sqrt { - g} \left( {\left[ {2{a_5} + 2n{a_6}} \right]{\nabla ^2}\ln \Omega } \right.  \notag \\
  &+ \left[ {4{a_7} + 4n{a_8} + 4{a_9} + 4n{a_{10}} + 4{n^2}{a_{11}}} \right]\left( {{\nabla ^\lambda }\ln \Omega } \right)\left( {{\nabla _\lambda }\ln \Omega } \right)  \notag \\
  &+ \left[ { - {a_5} - \left( {n - 2} \right){a_6} + 2{a_8} + 4{a_{10}} + 4n{a_{11}}} \right]{g^{\mu \nu }}{Q^\lambda }_{\mu \nu }{\nabla _\lambda }\ln \Omega   \notag \\
  &+ \left[ {2{a_5} + 4{a_7} + 2n{a_8} + 4{a_9}} \right]{Q^\mu }_{\mu \lambda }{\nabla ^\lambda }\ln \Omega   \notag \\
  &\left. { + \left[ {4{a_5} + 4n{a_6} + 2n{a_{12}} + 2{a_{13}} + 2{a_{14}}} \right]{S_{\lambda}}{\nabla ^\lambda }\ln \Omega } \right) \label{eq.13}
   \end{align}
in which ${\nabla ^2}\ln \Omega  = {\nabla _{\left\{ \lambda
\right\}}}{\nabla ^{\left\{ \lambda  \right\}}}\ln \Omega$. The
conformal invariance of the action results in vanishing the
additional terms at the above relation, that leads one to write:
\begin{align}
  & {a_6} =  - \frac{{{a_5}}}{n},  \notag \\
  & {a_9} =  - \frac{{{a_5}}}{2} - {a_7} - \frac{{n{a_8}}} {2},  \notag \\
  & {a_{11}} = \frac{{{a_5}}}{{2{n^2}}} - \frac{{{a_8}}} {{2n}} - \frac{{{a_{10}}}} {n},  \notag \\
  & {a_{14}} =  - n{a_{12}} - {a_{13}}. \label{copled eq}
\end{align}
Inserting (\ref{copled eq}) in (\ref{Mag action}) leads to
\begin{align}
{{\cal I}_{WMAG}} &= \frac{1} {\kappa}\int {{d^n}x\sqrt { - g}
\left( {{a_0}{\textbf{R}} + {a_1}{\nabla ^\mu }{S_{\mu}} +
{a_2}S_{\mu} S^{\mu}  + {a_3}{g^{\mu \nu }}{S_{\mu \lambda }}^\sigma
{S_{\nu \sigma }}^\lambda  + {a_4}{S_{\mu \nu \lambda}} {S^{\mu \nu
}}^\lambda }
\right.} \notag \\
&+ {a_5}\left[ {{g^{\mu \nu }}{\nabla _\mu }{Q^\sigma }_{\nu \sigma
} - \frac{1} {n}{g^{\nu \sigma }}{\nabla _\mu }{Q^\mu }_{\nu \sigma
} - \frac{1} {2}{g^{\mu \nu }}{Q^\sigma }_{\mu \lambda }{Q^\lambda
}_{\nu \sigma } + \frac{1} {{2{n^2}}}{g^{\mu \nu }}{g^{\alpha \beta
}}{g_{\lambda \gamma }}{Q^\lambda }_{\mu \nu }{Q^\gamma }_{\alpha
\beta }} \right] \notag \\
&+ {a_7}\left[ {{g^{\mu \nu }}{Q^\lambda }_{\mu \lambda }{Q^\sigma
}_{\nu \sigma }
- {g^{\mu \nu }}{Q^\sigma }_{\mu \lambda }{Q^\lambda }_{\nu \sigma }} \right] \notag \\
&+ {a_8}\left[ {{g^{\mu \lambda }}{Q^\nu }_{\mu \lambda }{Q^\sigma
}_{\nu \sigma } - \frac{n} {2}{g^{\mu \nu }}{Q^\sigma }_{\mu \lambda
}{Q^\lambda }_{\nu \sigma } - \frac{1} {{2n}}{g^{\mu \nu
}}{g^{\alpha \beta }}{g_{\lambda \gamma }}{Q^\lambda }_{\mu \nu
}{Q^\gamma }_{\alpha \beta }} \right] \notag \\
&+ {a_{10}}\left[ {{g^{\mu \alpha }}{g^{\nu \beta }}{g_{\lambda
\gamma }}{Q^\lambda }_{\mu \nu }{Q^\gamma }_{\alpha \beta } -
\frac{1} {n}{g^{\mu \nu }}{g^{\alpha \beta }}{g_{\lambda
\gamma}}{Q^\lambda
}_{\mu \nu }{Q^\gamma }_{\alpha \beta }} \right] \notag \\
&\left. { + {a_{12}}\left[ {{g^{\mu \nu }}{Q^\lambda }_{\mu \nu
}{S_{\lambda}} - n{g^{\mu \alpha }}{Q^\lambda }_{\mu
\nu }{S_{\lambda \alpha }}^\nu } \right] + {a_{13}}\left[ {{g^{\mu
\nu }}{Q^\lambda }_{\mu \lambda }{S_{\nu}} - {g^{\mu
\alpha }}{Q^\lambda }_{\mu \nu }{S_{\lambda \alpha }}^\nu } \right]}
\right). \label{W-action}
\end{align}
This is the general form of the Weyl-invariant metric-affine gravity
(WMAG), noting that a compensating Weyl scalar is needed to cancel
out the ${\Omega ^{n - 2}}$ factor which appears in (\ref{eq.13}).\\

More simplification can be done if one is interested in the
reduced form of metric-affine space, for example, in
Einstein-Weyl-Cartan space one has: ${\nabla _\lambda
}{g_{\mu \nu }} = - 2 {A_\lambda }{g_{\mu \nu }}$, where ${A_\lambda
}$ is a vector filed \cite{24}. By applying this condition
to (\ref{W-action}) it turns to the following simple form
\begin{equation}
{\cal I}_{WMAG} =\frac{1} {\kappa } \int d^n x\sqrt { - g} \Big(a_0
\textbf{R} + a_1\nabla^\mu S_\mu  + a_2 S_\mu S^\mu + a_3 g^{\mu
\nu} S_{\mu \lambda }\,^\sigma S_{\nu \sigma }\,^\lambda  + a_4
S_{\mu \nu \lambda} S^{\mu \nu\lambda}\Big)  \label{Weyl-Cartan}
\end{equation}

We would like to mention that (\ref{Weyl-Cartan}) is expressible in
a special form when the affine curvature scalar and covariant
derivative are redefined in terms of the Riemannian part plus a
particular combination of torsion and non-metricity square-terms by
using (\ref{affine connection}), (\ref{connection terms}) and
(\ref{affine curvature}). Therefore, the affine curvature scalar
takes the form of
\begin{align}
 \textbf{R} &= R(\{ \} ) - 4{g^{\mu \nu }}{\nabla _{\left\{ \mu  \right\}}}{S_{\nu}}
    - 4{S_{\mu}} {S^{\mu}}
    + 2{g^{\mu \nu }}{S_{\mu \lambda }}^\sigma {S_{\nu \sigma }}^\lambda
    + S_{\mu \nu\lambda} S^{\mu \nu\lambda}  \notag \\
   &+ {g^{\nu \sigma }}{\nabla _{\left\{ \mu  \right\}}}{Q^\mu }_{\nu \sigma }
   - {g^{\mu \nu }}{\nabla _{\left\{ \mu  \right\}}}{Q^\sigma }_{\nu \sigma } + \frac{1}
{2}{g^{\mu \lambda }}{Q^\nu }_{\mu \lambda }{Q^\sigma }_{\nu \sigma
} - \frac{1} {2}{g^{\mu \nu }}{Q^\sigma }_{\mu \lambda }{Q^\lambda
}_{\nu \sigma } + \frac{1} {4}{g^{\mu \alpha }}{g^{\nu \beta
}}{g_{\lambda \gamma }}{Q^\lambda }_{\mu \nu }{Q^\gamma }_{\alpha
\beta } \notag \\
   &- \frac{1}
{4}{g^{\mu \nu }}{g^{\alpha \beta }}{g_{\lambda \gamma }}{Q^\lambda
}_{\mu \nu }{Q^\gamma }_{\alpha \beta } + 2{g^{\mu \nu }}{Q^\lambda
}_{\mu \nu }{S_{\lambda }}  - 2{g^{\mu \nu }}{Q^\lambda }_{\mu
\lambda }{S_{\nu }}  - 2{g^{\mu \alpha }}{Q^\lambda }_{\mu \nu
}{S_{\lambda \alpha }}^\nu \label{affine curvature scalar}
\end{align}
and in Einstein-Weyl-Cartan space it becomes
\begin{multline}
  \textbf{R} = R(\{ \} )- 2\left( {n - 1} \right){\nabla _{\left\{ \sigma  \right\}}}{A^\sigma }
   - \left( {n - 1} \right)\left( {n - 2} \right){A^2} \\ - 4\left( {n - 2} \right){A^\mu }{S_{\mu}}
   - 4{g^{\mu \nu }}{\nabla _{\left\{ \mu  \right\}}}{S_{\nu}}
  - 4{S_{\mu}} {S^{\mu}}
  + 2{g^{\mu \nu }}{S_{\mu \lambda }}^\sigma {S_{\nu \sigma }}^\lambda
  + S_{\mu \nu\lambda} {S^{\mu \nu
  }}^\lambda. \label{redefineR}
\end{multline}
With the same procedure one obtains
\begin{equation}
{\nabla ^\mu }{S_{\mu}}  =
{g^{\mu \nu }}{\nabla _{\left\{ \mu  \right\}}}{S_{\nu }} + 2{S^{\mu }} {S_{\mu}} + \left( {n - 2} \right){A^\mu }{S_{\mu}}
  \label{redefinea1}
\end{equation}
and substituting (\ref{redefineR}) and (\ref{redefinea1}) in
(\ref{Weyl-Cartan}) results in
\begin{align}
{{\cal I}_{WMAG}^{sp.}} &= \frac{1} {\kappa }\smallint {d^n}x\sqrt {
- g} \left( {{a_0}R(\{ \} ) - 2\left( {n - 1} \right){a_0}{\nabla
_{\left\{ \sigma \right\}}}{A^\sigma } - \left( {n - 1}
\right)\left( {n - 2} \right){a_0}{A^2}} \right. - \left( {4{a_0} -
{a_1}} \right){g^{\mu \nu }}{\nabla _{\left\{ \mu  \right\}}}{S_{\nu}}
\notag\\
   &- \left( {4{a_0} - 2{a_1} - {a_2}} \right)S_{\mu}S^{\mu}
   + \left( {2{a_0} + {a_3}} \right){g^{\mu \nu }}{S_{\mu \lambda }}^\sigma {S_{\nu \sigma }}^\lambda
   + \left( {{a_0} + {a_4}} \right)S_{\mu \nu \lambda} S^{\mu \nu\lambda}
   \notag\\
  &\left. { - \left( {4\left( {n - 2} \right){a_0} - \left( {n - 2} \right){a_1}} \right){A^\mu }{S_{\mu}}} \right). \label{special action}
\end{align}
Now by imposing the torsion-free condition on (\ref{special action})
and setting $a_0=1$, up to a compensating Weyl scalar it becomes
similar to what studied as a conformally invariant extension of
Einstein-Hilbert action \cite{29}.\\

It is worth noting that when the action in (\ref{W-action}) is
transferred to Weitzenb\"{o}ck space by setting the curvature and
non-metricity to zero \cite{11}, it reduces to
\begin{equation}
{{\cal I}_{WMAG}} = \frac{1} {\kappa}\int {{d^n}x\sqrt { - g} }
\left( a_1 \nabla ^\mu S_{\mu}  +
a_2 S_{\mu}S^{\mu}  + a_3 g^{\mu \nu } S_{\mu \lambda }\,^\sigma S_{\nu
\sigma }\,^\lambda  + {a_4}S_{\mu \nu \lambda} S^{\mu\nu\lambda} \right)
\end{equation}
which is similar to the one used in teleparallel theories of gravity
\cite{18, 28}. For example, in Ref. \cite{conformaltele}, from the
tetrad and a scalar field analysis of torsion, it is shown that such
an action with specific coefficients of $a_1=0$, $a_2=-\frac{1}{3}$,
$a_3=\frac{1}{2}$ and $a_4=\frac{1}{4}$, can indeed be a conformally
invariant teleparallel action.

\section{Conclusion}

In this paper we have first studied the general form of second order
metric-affine action which is constructed from all the possible
forms of 15 scalar terms made up of affine curvature, torsion and
non-metricity tensors. The Weyl-invariant extension is obtained by
imposing conformal invariance as a condition on the action. The
resultant action reduces to the conformally invariant teleparallel
action in transition to Weitzenb\"{o}ck space \cite{conformaltele}.
Studying conformally invariant torsion theories are important
because aside from the conformal invariance property, they can
address some important issues of theoretical physics (for example
see \cite{pol}). Torsion theories may be viewed as a rank-3 mixed
symmetry tensor field. From the space-time symmetry and group
theoretical point of view, it is proved that linear conformal
quantum gravity in flat and de Sitter backgrounds should contain
such mixed symmetry tensor filed of rank-3 \cite{bine, jhep, jmp}.

Our obtained action (\ref{Mag action}) and its Weyl-invariant
extension covers the theories that made up from Riemann curvature
equipped with the torsion and non-metricity, albeit studying the
Weyl invariance of such theories needs some modifications, however
as discussed, in our method the Weyl invariance can only be obtained
by fixing the coefficients, that we have considered it as a special
case.

{\subsection*{Acknowledgement}

The author would like to thank M.V. Takook for his helpful
guidelines.

{\subsection*{Appendix A: Field equations of the general
action}\label{App. A}

\smallskip
In this appendix we obtain the related field equations of the
(\ref{Mag action}) in 4 dimensions. At first let us variate the
action with respect to the metric, which leads to:
\begin{align}
& {a_0}\left\{ {{\textbf{R}_{(\mu \nu )}} - \frac{1} {2}{g_{\mu \nu
}}\textbf{R}} \right\} + {a_1}\left\{ {{\nabla _{(\mu }}{S_{\nu )}}
- \frac{1} {2}{g_{\mu \nu }}{\nabla ^\beta }{S_\beta }} \right\} +
{a_2}\left\{ { - \frac{1} {2}{g_{\mu \nu }}{S_\alpha }{S^\alpha } +
{S_\mu }{S_\nu }} \right\}  \notag\\
  &  + {a_3}\left\{ { - \frac{1}
{2}{g_{\mu \nu }}{S_{\alpha \lambda \sigma }}{S^{\alpha \sigma
\lambda }} + {S_{\mu \lambda }}^\sigma {S_{\nu \sigma }}^\lambda }
\right\} + {a_4}\left\{ { - \frac{1} {2}{g_{\mu \nu }}{S_{\rho
\sigma \lambda }}{S^{\rho \sigma \lambda }} + 2{S^\alpha }_{\nu
\lambda }{S_{\alpha \mu }}^\lambda  - {S_{\rho \sigma \mu }}{S^{\rho
\sigma }}_\nu } \right\}  \notag\\
  &  + {a_5}\left\{ { - {\nabla _\nu }{\nabla _\mu }\ln \sqrt { - g}
   - \left( {{\nabla _\nu }\ln \sqrt { - g} } \right)\left( {{\nabla _\mu }\ln \sqrt { - g} } \right)
   + 2{Q^\alpha }{{_\alpha }_{(\mu }}{\nabla _{\nu )}}\ln \sqrt { - g}  + 4{S_{(\mu }}{\nabla _{\nu )}}\ln \sqrt { - g} } \right.  \notag\\
  & \left. { + {g_{\mu \kappa }}{\nabla _\nu }{Q_\alpha }^{\alpha \kappa } + {\nabla _\mu }{Q^\sigma }_{\sigma \nu } - \frac{1}
{2}{g_{\mu \nu }}{\nabla ^\beta }{Q^\sigma }_{\sigma \beta } +
2{\nabla _\nu }{S_\mu } + {Q_\rho }^{\rho \gamma }{Q_{\nu \gamma \mu
}} - {Q^\sigma }_{\sigma \nu }{Q^\kappa }_{\kappa \mu } - 4{S_\mu
}{S_\nu } - 4{S_{(\mu }}{Q^\sigma }_{\sigma |\nu )}} \right\}  \notag\\
  &  + {a_6}\left\{ { - {g_{\mu \nu }}{\nabla _\beta }{\nabla ^\beta }\ln \sqrt { - g}
   - {g_{\mu \nu }}\left( {{\nabla _\beta }\ln \sqrt { - g} } \right)\left( {{\nabla ^\beta }\ln \sqrt { - g} } \right)
    - {Q_{\nu \sigma }}^\sigma {\nabla _\mu }\ln \sqrt { - g}  + 2{Q^\alpha }_{\mu \nu }{\nabla _\alpha }\ln \sqrt { - g} } \right.  \notag\\
  &  + 4{g_{\mu \nu }}{S^\beta }{\nabla _\beta }\ln \sqrt { - g}  + 2{\nabla _\alpha }{Q^\alpha }_{\mu \nu } - \frac{1}
{2}{g_{\mu \nu }}{g^{\sigma \beta }}{\nabla _\alpha }{Q^\alpha
}_{\sigma \beta } + 2{g_{\mu \nu }}{\nabla _\beta }{S^\beta } +
{Q_\mu }^{\sigma \rho }{Q_{\nu \sigma \rho }} - 2{Q^{\alpha \lambda
}}_\nu {Q_\alpha }_{\lambda \mu }  \notag\\
  & \left. { - 4{g_{\mu \nu }}{S^\alpha }{S_\alpha } + 2{S_\mu }{Q_{\nu \sigma }}^\sigma  - 4{S_\alpha }{Q^\alpha }_{\mu \nu }} \right\}  \notag\\
  &  + {a_7}\left\{ {2{Q^\beta }_{\beta \mu }{\nabla _\nu }\ln \sqrt { - g}  + 2{\nabla _\nu }{Q^\beta }_{\beta \mu } - \frac{1}
{2}{g_{\mu \nu }}{Q^\lambda }_{\lambda \alpha }{Q_\sigma }^{\sigma
\alpha } - {Q^\beta }_{\beta \mu }{Q^\lambda }_{\lambda \nu } -
4{S_\nu }{Q^\beta }_{\beta \mu }} \right\}  \notag\\
  &  + {a_8}\left\{ {{Q_{\mu \beta }}^\beta {\nabla _\nu }\ln \sqrt { - g}
  + {g_{\mu \nu }}{Q_\beta }^{\beta \sigma }{\nabla _\sigma }\ln \sqrt { - g}
  + {g_{\mu \nu }}{\nabla ^\kappa }{Q^\beta }_{\beta \kappa } + {g^{\beta \lambda }}{\nabla _\nu }{Q_{\mu \beta \lambda }}
   + 2{Q^\sigma }_{\sigma [\mu }{Q_{\nu ]\beta }}^\beta } \right. - {Q_\nu }^{\beta \lambda }{Q_{\mu \beta \lambda }}  \notag\\
  & \left. { - {g_{\mu \nu }}{Q_\sigma }^{\sigma \kappa }{Q^\beta }_{\beta \kappa } - \frac{1}
{2}{g_{\mu \nu }}{Q^{\beta \lambda }}_\lambda {Q^\sigma }_{\sigma
\beta } - 2{S_\nu }{Q_{\mu \sigma }}^\sigma  - 2{g_{\mu \nu
}}{S^\rho }{Q^\beta }_{\beta \rho }} \right\}  \notag\\
  &  + {a_9}\left\{ {2{Q_{\nu \mu \sigma }}{\nabla ^\sigma }\ln \sqrt { - g}  - 2{Q_{\nu \mu \sigma }}{Q_\beta }^{\beta \sigma }
   + 2{\nabla ^\sigma }{Q_{\nu \mu \sigma }} - \frac{1}
{2}{g_{\mu \nu }}{Q^{\sigma \beta \lambda }}{Q_{\lambda \beta \sigma
}} - {Q^\lambda }_{\mu \sigma }{Q^\sigma }_{\lambda \nu } -
4{S^\sigma }{Q_{\nu \mu \sigma }}} \right\}  \notag\\
  &  + {a_{10}}\left\{ {2{Q^\sigma }_{\mu \nu }{\nabla _\sigma }\ln \sqrt { - g}  + 2{\nabla _\sigma }{Q^\sigma }_{\mu \nu } - \frac{1}
{2}{g_{\mu \nu }}{Q^{\lambda \alpha \beta }}{Q_{\lambda \alpha \beta
}} + {Q_\nu }^{\alpha \beta }{Q_{\mu \alpha \beta }} - 2{Q^{\sigma
\alpha }}_\nu {Q_{\sigma \alpha \mu }} - 4{S_\sigma }{Q^\sigma
}_{\mu \nu }} \right\}  \notag\\
  &  + {a_{11}}\left\{ {2{g_{\mu \nu }}{Q_{\sigma \beta }}^\beta {\nabla ^\sigma }\ln \sqrt { - g}
   + 2{g_{\mu \nu }}{g^{\alpha \beta }}{\nabla _\sigma }{Q^\sigma }_{\alpha \beta } - \frac{1}
{2}{g_{\mu \nu }}{Q^{\lambda \sigma }}_\sigma {Q_{\lambda \alpha
}}^\alpha  + {Q_{\mu \alpha }}^\alpha {Q_{\nu \rho }}^\rho  -
2{g_{\mu \nu }}{Q^{\sigma \alpha \beta }}{Q_{\sigma \alpha \beta }}}
\right.\left. { - 4{g_{\mu \nu }}{S^\sigma }{Q_{\sigma \beta
}}^\beta } \right\}  \notag\\
  &  + {a_{12}}\left\{ {{g_{\mu \nu }}{S^\beta }{\nabla _\beta }\ln \sqrt { - g}  + {g_{\mu \nu }}{\nabla _\beta }{S^\beta }
  - 2{g_{\mu \nu }}{S^\lambda }{S_\lambda } - \frac{1}
{2}{g_{\mu \nu }}{S^\lambda }{Q_{\lambda \beta }}^\beta  + {S_\mu
}{Q_{\nu \alpha }}^\alpha } \right\}  \notag\\
  &  + {a_{13}}\left\{ {{S_\mu }{\nabla _\nu }\ln \sqrt { - g}  + {\nabla _\nu }{S_\mu } - 2{S_\nu }{S_\mu } - \frac{1}
{2}{g_{\mu \nu }}{S^\alpha }{Q^\lambda }_{\lambda \alpha } +
2{Q^\lambda }_{\lambda [\mu }{S_{\nu ]}}} \right\}  \notag\\
  &  + {a_{14}}\left\{ {{S_{\lambda \mu \nu }}{\nabla ^\lambda }\ln \sqrt { - g}  + {g_{\rho \nu }}{\nabla ^\lambda }{S_{\lambda \mu }}^\rho
   - 2{S^\lambda }{S_{\lambda \mu \nu }} - \frac{1}
{2}{g_{\mu \nu }}{S^{\lambda \beta \rho }}{Q_{\lambda \beta \rho }}
- {S_{\lambda \alpha \nu }}{Q^{\lambda \alpha }}_\mu } \right.  \notag\\
  & \left. { + {S_\mu }^{\beta \rho }{Q_{\nu \beta \rho }} - {S_{\lambda \mu \nu }}{Q_\sigma }^{\sigma \lambda }
   + {S_{\lambda \nu }}^\beta {Q^\lambda }_{\mu \beta }} \right\} =  - \frac{\kappa}
{{\sqrt { - g} }}\frac{{\delta {{\cal I}_{M}} }} {{\delta {g^{\mu
\nu }}}}
\end{align}
where ${{\cal I}_{M}} $ stands for matter action.

Variation of the action with respect to the connection yields the
following equation:
\begin{align}
  & {a_0}\left\{ { - {g^{\mu \nu }}{\nabla _\lambda }\ln \sqrt { - g}  + {\delta ^\nu }_\lambda {\nabla ^\mu }\ln \sqrt { - g}  + {Q_\lambda }^{\mu \nu } - {Q_\sigma }^{\sigma \mu }{\delta ^\nu }_\lambda  - 2{S_\lambda }^{\mu \nu } + 2{g^{\mu \nu }}{S_\lambda } - 2{S^\mu }{\delta ^\nu }_\lambda } \right\}  \notag\\
  &  + {a_1}\left\{ {{\delta ^{[\mu }}_\lambda {\nabla ^{\nu ]}}\ln \sqrt { - g}  + {Q_\alpha }^{\alpha [\mu }{\delta ^{\nu ]}}_\lambda  - {g^{\mu \nu }}{S_\lambda } + 2{S^{[\mu }}{\delta ^{\nu ]}}_\lambda } \right\} + {a_2}\left\{ {2{S^{[\mu }}{\delta ^{\nu ]}}_\lambda } \right\} + {a_3}\left\{ {2{S_\lambda }^{\left[ {\nu \mu } \right]}} \right\}  \notag\\
  &  + {a_4}\left\{ {2{S^{\left[ {\mu \nu } \right]}}_\lambda } \right\} + {a_5}\left\{ {{g^{\mu \nu }}{\nabla _\lambda }\ln \sqrt { - g}  + \delta _\lambda ^\nu {\nabla ^\mu }\ln \sqrt { - g}  - 2{g^{\mu \nu }}{Q^\gamma }_{\gamma \lambda } - {Q_\rho }^{\rho \mu }\delta _\lambda ^\nu  - 2{g^{\mu \nu }}{S_\lambda } - 2\delta _\lambda ^\nu {S^\mu }} \right\}  \notag\\
  &  + {a_6}\left\{ { 2\delta _\lambda ^\mu {\nabla ^\nu }\ln \sqrt { - g}  - 4{Q^{\nu \mu }}_\lambda  + {Q^\mu }{{^\gamma }_\gamma }\delta _\lambda ^\nu  - 4{S^\nu }\delta _\lambda ^\mu } \right\}{\text{ + }}{a_7}\left\{ { - 2{g^{\mu \nu }}{Q^\beta }_{\beta \lambda } - 2{Q_\beta }^{\beta \mu }\delta _\lambda ^\nu } \right\}  \notag\\
  &  + {a_8}\left\{ { - {g^{\mu \nu }}{Q_{\lambda \rho }}^\rho  - 2{Q_\beta }^{\beta \nu }\delta _\lambda ^\mu  - {Q^{\mu \beta }}_\beta \delta _\lambda ^\nu } \right\} + {a_9}\left\{ { - 2{Q^{\mu \nu }}_\lambda  - 2{Q_\lambda }^{\mu \nu }} \right\} + {a_{10}}\left\{ { - 4{Q^{\nu \mu }}_\lambda } \right\}  \notag\\
  &  + {a_{11}}\left\{ { - 4{Q^{\nu \beta }}_\beta \delta _\lambda ^\mu } \right\} + {a_{12}}\left\{ {\delta _\lambda ^{[\nu }{Q^{\mu ]\beta }}_\beta  - 2\delta _\lambda ^\mu {S^\nu }} \right\} + {a_{13}}\left\{ {{Q_\alpha }^{\alpha [\mu }\delta _\lambda ^{\nu ]} - {g^{\mu \nu }}{S_\lambda } - {S^\mu }{\delta ^\nu }_\lambda } \right\}  \notag\\
  &  + {a_{14}}\left\{ {{Q^{[\mu \nu ]}}_\lambda  + {S^{\mu \nu }}_\lambda  + {S_\lambda }^{\nu \mu }} \right\} =  - \frac{\kappa}
{{\sqrt { - g} }}\frac{{\delta {{\cal I}_{M}} }} {{\delta {\Gamma
^\lambda}_{\mu \nu }}}.
\end{align}



\begin{thebibliography}{}
\bibliographystyle{unsrt}

\bibitem{1} S. Capozziello, V. Faraoni, \emph{Beyond Einstein Gravity: A Survey of Gravitational Theories for Cosmology and Astrophysics}, Springer (2011).

\bibitem{2} S. Nojiri, S.D. Odintsov, \textit{Introduction to Modified Gravity and Gravitational Alternative for Dark Energy}, Int. J. Geom. Meth. Mod. Phys. 4 (2007) 115-146, arXiv:hep-th/0601213;\\
S. Capozziello, M. De Laurentis, \emph{Extended Theories of Gravity}, Phys. Rep. 509, Issues 4-5,  (2011), 167,  arXiv: 1108.6266v2.

\bibitem{3} H. F. M. Goenner, \emph{On the History of Unified Field Theories}, Living Rev. Relativity, 7, (2004), 2. [Online Article]: cited, \url{http://www.livingreviews.org/lrr-2004-2}.

\bibitem{3.1} S. Nojiri, S. D. Odintsov, \textit{Unified cosmic history in modified gravity: from F(R) theory to Lorentz non-invariant models},  Phys. Rept. 505 (2011) 59-144, arXiv:1011.0544 [gr-qc].

\bibitem{hehl} F. W. Hehl, J. D. McCrea, E. W. Mielke and Y. Neeman, \textit{Metric-affine gauge theory of gravity: field equations, Noether identities, world spinors, and breaking of dilation invariance}, Phys. Rep. 258, Issues 1-2, (1995), Pages 1-171;\\
 D. Puetzfeld, \emph{Prospects of non-Riemannian cosmology}, Proceedings of 22nd Texas Symposium on Relativistic Astrophysics at Stanford University, Stanford, California, 13-17 Dec (2004), arXiv: astro-ph/0501231v1;\\
 D. Puetzfeld, \emph{Status of non-Riemannian cosmology}, New Astron. Rev. 49 (2005) 59;\\
 A. V. Minkevich, A. S. Garkun, \emph{Isotropic cosmology in metric-affine gauge theory of gravity}, arXiv: gr-qc/9805007v1;\\
  N. J. Poplawski, \emph{Acceleration of the universe in the Einstein frame of a Metric-affine f(R) gravity}, Class. Quant. Grav. 23 (2006) 2011;\\
  D. Puetzfeld, Y. N. Obukhov, \emph{Probing non-Riemannian spacetime geometry}, Phys. Lett. A 372, (2008) 6711, arXiv: 0708.1926v3.

\bibitem{5} F. W. Hehl, Y. N. Obukhov, \emph{Elie Cartan's torsion in geometry and in field theory, an essay}, arXiv:0711.1535.

\bibitem{11} Yi Mao, M. Tegmark, A. Guth, S. Cabi, \emph{Constraining Torsion with Gravity Probe B}, Phys. Rev. D 76, (2007) 104029, arXiv:gr-qc/0608121v4.

\bibitem{stelle} K. S. Stelle, \textit{Renormalization of higher-derivative quantum gravity}, Phys. Rev. D 16, (1977) 953.

\bibitem{mann} P. D. Mannheim and D. Kazanas, \textit{Exact vacuum solution to conformal Weyl gravity and galactic rotation curves}, Astrophys. J. 342, (1989) 635;\\
P. D. Mannheim, \textit{Conformal cosmology with no cosmological constant}, Gen. Rel. Grav. 22, (1990) 289.

\bibitem{luca} L. Fabbri,  \emph{Metric-Torsional Conformal Gravity}, Phys. Lett. B 707, (2012) 415, arXiv: 1101.1761v3.
\\There are
other choices of the conformal transformation of the torsion tensor
which are studied in \cite{OBUKHOV, J. W. Maluf, 23, 24}.

\bibitem{turk} C. N. Karahan, O. Dogangun, D. A. Demir, \emph{Conformal Transformations in Metric-Affine Gravity and Ghosts},  Ann. Phys. (Berlin) 524 (2012) 461, arXiv:1204.6366v2.

\bibitem{13} V. Vitagliano, T. P. Sotiriou, S. Liberati, \emph{The dynamics of metric-affine gravity}. Annals Phys. 326 (2011) 1259, arXiv: 1008.0171v3.

\bibitem{14} T. P. Sotiriou, \emph{Modified Actions for Gravity: Theory and Phenomenology}, International School for Advanced Studies, (2007), arXiv:0710.4438.

\bibitem{15} M. Blau, \emph{Lecture Notes on General Relativity}, Albert Einstein Center for Fundamental Physics, Bern University, \url{http://www.blau.itp.unibe.ch/Lecturenotes.html}, (2012).

\bibitem{16} A. Kleyn, \emph{Metric-Affine Manifold}, arXiv:gr-qc/0405028.

\bibitem{17} T. P. Sotiriou, \emph{The significance of matter coupling in f(R) gravity}, Proceedings of the Eleventh Marcel Grossmann Meeting on General Relativity, World Scientific, Singapore (2008), arXiv:gr-qc/0611158v1.

\bibitem{note} In our study only the geometrical part of the action is dealt with, so the inconsistency that is mentioned in \cite{Hehl,18}, does not appear in our consideration.

 \bibitem{Hehl} F. W. Hehl,  G.D. Kerlick, P. Von der Heyde \textit{On a new metric affine theory of gravitation}, Phys. Lett. B, Volume 63, Issue 4, (1976), 446.

\bibitem{18} T. Clifton, P. G. Ferreira, A. Padilla, C. Skordis, \emph{Modified Gravity and Cosmology}, Phys. Rep. 513, 1 (2012) 1-189, arXiv:1106.2476v3.

\bibitem{19} V. Vitagliano,  \emph{The role of nonmetricity in metric-affine theories of gravity}, arXiv:1308.1642v1.

\bibitem{10} C. N. Karahan, A. Altas, D. A. Demir, \emph{Scalars, Vectors and Tensors from Metric-Affine Gravity}, Gen. Rel. Grav. 45 (2013) 319, arXiv: 1110.5168v1.

\bibitem{21} T. Fulton, F. Rohrlich, L. Witten, \emph{Conformal Invariance in Physics}, Rev. Mod. Phys. Vol. 34 no. 3 (1962);\\L. O'Raifeartaigh, I. Sachs and C. Wiesendanger, \emph{Weyl gauging and curved space approach to scale and conformal invariance}, Meeting on 70 Years of Quantum Mechanics, Calcutta, India, 29 Jan - 2 Feb (1996).

\bibitem{OBUKHOV} Y. N. Obukhov, \emph{Conformal invariance and space-time torsion}, Physics Letters A, Volume 90, Issue 1-2, p. 13-16 (1982).

\bibitem{Kazuharu} K. Bamba, S. D. Odintsov, D. Sáez-Gómez, \textit{Conformal symmetry and accelerating cosmology in teleparallel gravity}, Phys. Rev. D 88, (2013) 084042, arXiv:1308.5789 [gr-qc].

\bibitem{J. W. Maluf} J. W. Maluf, \emph{Conformal invariance and torsion in general relativity}, Gen. Rel. Grav. 19 (1987) 57.

\bibitem{23} I. L. Shapiro, \emph{Physical Aspects of the Space-Time Torsion}, Phys. Rept. 357 (2002) 113, arXiv: hep-th/0103093v1.

\bibitem{24} T. Y. Moon, J. Lee, Phillial Oh, \emph{Conformal Invariance in Einstein-Cartan-Weyl space},  Mod. Phys. Lett. A 25, (2010) 3129 , arXiv:0912.0432v3.

\bibitem{29} M. P. Dabrowski and J.
Garecki, \emph{Conformal transformations and conformal invariance in
gravitation}, Annalen Phys. (Berlin) 18 (2009), 13-32,
arXiv:0806.2683.

\bibitem{28} R. Aldrovandi, J. G. Pereira, \emph{Gravitation: in search of the missing torsion}, Ann. Fond. Louis de Broglie Vol. 32 (2007) 229, arXiv: 0801.4148v1.

\bibitem{conformaltele} J. W. Maluf, F. F. Faria, \emph{Conformally invariant teleparallel theories of gravity},  Phys. Rev. D 85, (2012) 027502, arXiv:1110.3095 [gr-qc].

\bibitem{pol} N. J. Poplawski, \emph{Matter-antimatter asymmetry and dark matter from torsion}, Phys. Rev. D 83, (2011) 084033, arXiv:1101.4012 [gr-qc]; \emph{Nonsingular, big-bounce cosmology from spinor-torsion coupling},  Phys. Rev. D 85, (2012) 107502, arXiv:1111.4595 [gr-qc]; \emph {Spacetime torsion as a possible remedy to major problems in gravity and cosmology}, arXiv:1106.4859 [gr-qc].

\bibitem{bine} B. Binegar, C. Fronsdal, and W. Heidenreich, \emph{Linear conformal quantum gravity}, Phys. Rev. D 27, (1983) 2249.

\bibitem{jhep} M.V. Takook and M.R. Tanhayi, \emph{Linear Weyl Gravity in de Sitter Universe}, JHEP 1012: 044, (2010),  arXiv:0903.2670.

\bibitem{jmp} M.V. Takook, M.R. Tanhayi and S. Fatemi, \emph{Conformal linear gravity in de Sitter space},  J. Math. Phys. 51, (2010) 032503,  arXiv:0903.5249.

\end{thebibliography}
\end{document}